# Effects of Fission-Fragment Damage on Vortex Dimensionality in Silver-sheathed $Bi_2Sr_2Ca_2Cu_3O_x$ Tapes.

D. Marinaro*, J. Horvat and S. X. Dou
*Institute for Superconducting and Electronic Materials, University of Wollongong, Wollongong, NSW 2522, Australia*
R. Weinstein and A. Gandini
*Texas Center for Superconductivity, University of Houston, Houston, TX 77204-5506, USA.*

*Abstract*

We report on the vortex dimensionality of uranium-doped Ag/Bi2223 tapes, before and after irradiation to a thermal-neutron fluence. The effective activation energies as a function of current density and applied field were calculated from dynamic magnetisation relaxation measurements. A dimensional crossover from a 3D elastic creep regime to 2D plastic creep was observed in the nonirradiated tape at an applied magnetic field $\mu_0 H_{cr} \approx 0.37$ T, with an associated change in the flux hop velocity and temperature dependence. After introduction of the fission-fragment damage by irradiation, a shift in the crossover to $\mu_0 H_{cr} \approx 0.65$ T was observed. These results indicate an enhancement of the *c*-axis vortex coherence by the introduced splayed columnar defects, explaining the greater pinning efficiency of the uranium-fission method in Bi2223 rather than the less anisotropic Y123. Conflicting results obtained for the irradiated tape in the absence of any temperature scaling of the activation energies demonstrate the importance of the inclusion of scaling in the magnetisation relaxation analysis.

## I. INTRODUCTION

The nature of the vortex phase diagram in High-Temperature Superconductors (HTS) has been the subject of intense investigation in the 15 years since their discovery. Due to the combination of their high critical temperature, short coherence length and large anisotropy, HTS have a greater dependence on thermal fluctuations than that of conventional superconductors, resulting in vortex dynamics that are largely different and a more complex *H-T* phase diagram structure.

The large anisotropy in superconductor properties is a result of the strongly layered structure of oxide HTS. Combined with the short coherence lengths, a de-coupling transition from 3-dimensional (3D) vortex lines to 2D vortex pancakes is predicted[1]. The transition in vortex dimensionality results from a loss of the inter-layer vortex pancake coupling due to the influence of thermal and magnetic fluctuations on the Josephson and electromagnetic coupling. Evidence of this transition has been experimentally observed in $YBa_2Cu_3O_x$,[2] (Y123) and $Bi_2Sr_2CaCu_2O_x$[3,4] (Bi2212) systems. A thermally induced crossover has been observed in Ag-sheathed $Bi_2Sr_2Ca_2Cu_3O_x$ (Ag/Bi2223) tapes via dc flux transformer measurements[5] and magnetic relaxation measurements.[6]

The controlled creation of pinning centres allowed by heavy-ion irradiation techniques provides a suitable method for probing the vortex dynamics in a systematic manner.[7] There is mounting evidence that parallel columnar defects created across the layers of anisotropic HTS increase the inter-layer vortex coherence, promoting a 3D line-like behaviour, by suppressing the effects of thermal fluctuations on the pancake positions.[8] This effect has been observed in heavy-ion irradiation studies on Bi2212 single crystals[9] and thin films,[3] $Tl_2Ba_2CaCu_2O_x$ (Tl2212) thin films[10] and BSCCO tapes,[11] using diverse methods such as Josephson plasma resonance,[9] resistive dissipation[10] and $J_c(\theta)$ scaling.[11] The studies of transport $J_c(\theta)$ scaling with $B_\perp$,[11] the component of applied field perpendicular to the tape ab-plane, identified an increase in the 3D line-like vortex state in both Ag/Bi2223 and Ag/Bi2212 tapes with parallel columnar defects.

The introduction of splay (dispersion) in the columnar track directions was proposed[12] to improve pinning further, by suppressing the relaxation via double-kink formation that is the dominant type of vortex excitation in the parallel track configuration. Studies in Y123[13], however, demonstrated that sufficiently large splay angles would instead promote vortex motion and stimulate creep. This behaviour of large splay was explained by the action of vortex cutting,[14,15] whereby the *c*-axis coherence cannot be maintained, reducing the pinning action to that of point defects.[15]

Further, comparisons of the uranium-fission method on Y123, Bi2212 and Bi2223 samples revealed that pinning enhancements were greater in the more highly anisotropic BSCCO superconductors, particularly Bi2223.[16] At the time, the authors were unable to provide any structural evidence to explain the differences. Krusin-Elbaum *et al.*[17] investigated the effects of randomly splayed fission-induced pinning in Hg cuprates with different degrees of anisotropy. The authors demonstrated that uniformly splayed defects provide much stronger improvement of vortex pinning in superconductors with a high anisotropy and suggested that this results from a re-scaling of the pinning distribution by the anisotropy.

Based on the findings in Ref. 17, we would expect that randomly splayed columnar pinning tracks in Bi2223 would produce similar results to fission-damage in the most anisotropic of Hg cuprates. That is, we expect firstly a strong suppression of the flux creep rate (and hence a large increase in the activation energy barriers to flux creep) and secondly promotion of the $c$-axis vortex coherence in comparison to virgin Bi2223.

Enhancements in the inter-layer vortex coherence by heavy-ion irradiation induced parallel columnar defects have been reported in Ag/Bi2223 tapes.[11] The effect of random splay on the vortex coherence in Bi2223, however, has not been experimentally resolved.

We use a uranium-fission technique to introduce randomly splayed, quasi-columnar pinning defects within the superconducting matrix of highly $c$-axis oriented Ag/Bi2223 tapes. The effective activation energy $U_{eff}$ of the tapes before and after uranium fission was determined via dynamic magnetisation relaxation, and the vortex dimensionality was deduced from the behaviour of $U_{eff}$ as a function of current density.

In the course of this investigation, the necessity for the introduction of a temperature scaling term into the calculation of $U_{eff}(J)$ was made apparent. The best method for determining the functional form of this scaling term is empirically from the data, rather than from a form determined *a priori*. However, a conflicting result was determined in the absence of any temperature scaling of the effective activation energy. The outcomes of the analysis of the data without temperature scaling would imply a reduction in the $c$-axis vortex coherence after introduction of randomly splayed pinning, in direct contrast to the expectations based on the results of Ref. 17. The presence of a temperature scaling term provides a more physically reasonable value for the flux hop velocity. Combined with agreement from transport voltage – current measurements, we therefore conclude that the temperature-scaled analysis indicating promotion of the vortex coherence by the splayed defects is the correct interpretation.

A dimensional crossover from a 3D elastic creep regime to 2D plastic creep was observed in the virgin tape at an applied magnetic field $\mu_0 H_{cr} \approx 0.37$ T with an associated change in the flux hop velocity and the temperature dependence of the effective activation energy. After introduction of the fission-fragment damage, the crossover was shifted to $\mu_0 H_{cr} \approx 0.65$ T, indicating an enhancement of the $c$-axis vortex correlation due to the introduced splayed columnar defects.

## II. ANALYSIS THEORY

Measurements of the magnetisation relaxation provide a useful means of investigating the vortex dynamics. The effective activation energy, $U_{eff}$, for thermally activated flux hopping out of a pinning centre can be examined over a wide range of current densities by analysing the relaxation at various temperatures and applied fields. Theories of flux dynamics, such as collective flux creep[18] or vortex-glass theories,[19] predict the behaviour of $U_{eff}$ as a function of current density, so that the observed current dependence can be related to the dominant vortex dynamics.

There are a number of different methods available for analysis of the relaxation data.[20] One of the most widely used, and the method employed here, is Maley's method.[21] Dynamic and static relaxation measurements have been shown equivalent,[22] so we applied Maley's method after adjustment for dynamic magnetic relaxation data.[22] In considering thermally activated flux motion, the effective activation energy required for a flux hop out of a pinning centre can be described by

$$U_{eff}(J,H,T) = k_B T \left[ \ln\left(\frac{H}{\dot{H}}\right) + C \right]; \; C = \ln\left(\frac{2av_0}{d}\right), \quad (1)$$

where $a$ is the flux hop distance, $v_0$ is an attempt frequency, $d$ is a characteristic sample dimension, typically the average grain diameter and $\dot{H}$ is the applied field sweep rate in T/s. $C$ is a fitting parameter which is adjusted to provide a smooth, continuous $U_{eff}(J)$ curve, and is assumed to be temperature independent at $T \ll T_c$ (or only weakly temperature dependent at higher $T$). This assumption stems from the reasoning that $C$ does not have a strong dependence on $a$ and $v_0$, both temperature dependent terms, as they are only logarithmic arguments in $C$.[23] The resulting data points can then be fitted using the interpolation formula,[24]

$$U_{eff}(J,H,T) = \frac{U_c(H)}{\mu}\left[\left(\frac{J_c(T)}{J}\right)^\mu - 1\right], \quad (2)$$

where $U_c$, $J_c(T)$ and $\mu$ are fitting parameters, and $\mu$ is dependent on the flux dynamics. For $J \ll J_c(T)$, Eq. (2) reduces to the form predicted by the vortex-glass and collective creep theories,[18,19] $U \propto [J_c(T)/J]^\mu$, with $\mu > 0$.

The size and variation of $C$ with applied field required for a smooth fit of the data over a large temperature range was perceived to be physically unreasonable[23]. To overcome this, the introduction of a temperature scaling term was suggested, to separate the explicit temperature dependence of $U_{eff}$ from the current dependence. This *modified* Maley's method can be verified by considering $U_{eff}(J,H,T)$ as a separable function[25] of applied field, temperature and current density. Then $U_{eff}(J,H,T)$ can be expressed as

$$U_{eff}(J,H,T) = U_0 U(T) U(H) U(J), \qquad (3)$$

where $U_0$ is a characteristic pinning potential, and $U(T)$, $U(H)$ and $U(J)$ describe the temperature, field and current dependencies of $U_{eff}(J,H,T)$, respectively. It follows that

$$U_{eff}(J,H,T_0) = U_0 U(H) U(J) = \frac{U_{eff}(J,H,T)}{U(T)} \qquad (4)$$

is the effective activation energy scaled to a characteristic temperature $T_0$, and $U(H)$, $U(J)$ describe the field and current dependencies at $T_0$, respectively.

From the definition of $J_c = J$ at $U_{eff} = 0$ and with $U_{eff}(J,H,T_0)$ as defined in (4), Eq. (2) then becomes

$$U_{eff}(J,H,T_0) = \frac{U_c}{\mu}\left[\left(\frac{J_c(T_0)}{J}\right)^\mu - 1\right]. \qquad (5)$$

To determine $U(T)$, two different approaches have been taken. The first, as followed in Ref. 23, takes known temperature-scaling forms determined theoretically and applies them directly to the data. $U_{eff}(J,H,T)$ is divided by the scaling term and values for $C$ are chosen to provide the best continuity of $U_{eff}(J,H,T_0)$ over the entire temperature range. The scaling term can be changed or adjusted until a reasonable value of $C$ is found to provide a smooth continuous fit.

Tinkham,[26] by considering the temperature dependencies of $H_c$ and $\lambda$ in the Ginzburg-Landau relations, found a temperature scaling factor of the form

$$U(T) = 1 - t^2 (1-t^4)^{1/2}/t \; ; \; t = \frac{T}{T_c}, \qquad (6)$$

which can be approximated to

$$U(T) \approx (1-t)^{3/2}, \qquad (7)$$

valid over a wide range of temperatures above $t \approx \frac{1}{2}$. More specifically, this $U(T)$ reflects the temperature dependence of the depairing current density $J_{c0}$, which is $J_c$ at $H = 0$ and without any reductions by thermal fluctuations, and is thus by definition not reduced to zero above $T_{irr}$.[26] The form of Eq. (6) also predicts a form

$$U(T) \approx (1-t^2), \qquad (8)$$

at low temperatures, $t \ll \frac{1}{2}$. Temperature scaling functions of the form (7) and (8) were used in the literature to obtain smooth fit for U(J) in modified Maley's method. [23, 27]

The second approach involves determining $U(T)$ empirically from the experimental data.[27] The value of $C$ is chosen to provide continuity of $U_{eff}$ at the lowest isotherm data sets (highest $J$), where $C$ is expected to be the least temperature dependent. Each successive isotherm is adjusted by a multiplicative scaling factor $G$ that produces a continuous fit to the adjacent lower isotherms. Once all isotherms have been adjusted, the resulting curve should be a continuous $U_{eff}(J,H,T_0)$ curve. By curve fitting $G$ against $T$, the functional form can be determined, generally in the form[27]

$$G(T) = \frac{(1-t^m)^n}{a} = U(T)^{-1}. \qquad (9)$$

Measurement of the current-voltage characteristics over a range of applied fields at constant temperature provides a crosscheck for the field dependence of the sign of μ. By combining (1) and (2), and with the sweep rate $dH/dt \sim E$, one can derive[28]

$$\mu = -Q(T)\left[\frac{d^2(\ln E)}{(d\ln J)^2}\right]. \tag{10}$$

where $Q(T)$ is the dynamic relaxation rate. Thus, the parameter μ may be related to the curvature of the $\ln E$ vs $\ln J$ curves. A change in the sign of the curvature of $\ln E - \ln J$ as a function of magnetic field can therefore be taken as an indication of a similar change in the sign of μ, independent of any temperature scaling issues. Measurement of the current-voltage characteristics in various applied fields at a constant temperature provides a crosscheck for the field dependence of the sign of μ.

## III. EXPERIMENTAL METHODS

A mono-core Ag/Bi2223 tape doped with small quantities of $^{235}$U was fabricated using the powder-in-tube process.[29] The HTS precursor powder was doped with $c = 0.6$ wt% of $UO_2.2H_2O$. A highly enriched uranium mixture containing an isotopic ratio of approximately 98% $^{235}$U was used. $T_c$ was determined from ac susceptibility measurements to be 108K, and $J_{c0}$ from transport measurements was obtained as 22.4 kAcm$^{-2}$.

A 3 cm length of processed tape was irradiated by highly moderated thermal neutrons at the HIFAR reactor of the Australian Nuclear Science and Technology Organisation, to a fluence $\Phi_n = 2.25 \times 10^{19}$ m$^{-2}$. A second 3 cm length of tape was kept without irradiation as a control sample. The irradiation procedure resulted in a 1K reduction in $T_c$, and $J_{c0}$ was reduced to 18.1 kAcm$^{-2}$. The tapes are expressed in the form $(c, \Phi_n)$.

The thermal neutron fluence induces fission of the $^{235}$U, resulting in two high-energy heavy-ion fission-fragments with a mean mass of 115 amu and mean energy of about 85 MeV. The ionisation energy loss per unit length of the fission-fragments, $dE/dx$, varies along the fragment track, only occasionally exceeding the threshold value for amorphous track formation as a result of the stochastic nature of the process. The fission-fragment damage thus forms quasi-columnar defect tracks in a string-of-beads effect, with an average radius of 5 nm and average length of 10 μm per fission-fragment. The total angular distribution of the energetic fission-fragments is random, owing to the statistical nature of the fission process, so that the end result of the doping and irradiation procedure is the creation of randomly oriented, quasi-columnar amorphous defect tracks.

dc magnetic measurements were performed in an Oxford Instruments vibrating sample magnetometer on two sample pieces cut from the tapes measuring 4 x 3 x 0.2 mm$^3$. After zero-field cooling the sample to the measurement temperature, hysteresis loops were recorded in fields $\mu_0 H = \pm 1$ T, applied perpendicular to the tape plane. The hysteresis loops were measured at sweep rates of field of 0.05, 0.1, 0.25. 0.5, 0.75 and 1.0 T/min and within a temperature range 15 – 70 K. The current density was estimated from the irreversible magnetisation $M_{irr}$, half the difference between the magnetisation on the increasing and decreasing branches of the loops.

Transport voltage – current $(V – I)$ characteristics were measured using a standard four-probe dc method on the 2 cm lengths of tape. The samples were immersed in liquid nitrogen to reduce heating effects. Magnetic fields up to 0.6T were applied perpendicular to the tape plane. Voltages were pre-amplified by a Keithley 1801 nanovolt preamplifier and measured on a Keithley 2001 digital multimeter.

## IV. RESULTS

Applying the standard Maley's method to our experimental data, it was difficult to find a single value of $C$ that could provide smooth continuity of $U_{eff}$ over the entire temperature range. To line up isotherms in the low temperature range, values of $C \sim 15$ were used. However, this left the higher temperature isotherms extremely misaligned. To compromise in the quality of fit between the high and low temperature ranges, values of $C \sim 40$ were needed at low applied magnetic fields, and $C \sim 30$ at higher fields.

From (1), $2v_0/d \approx e^C$, where $v_0 \approx av_0$ is the flux hop velocity.[30] From the literature, values of $C$ between 10 and 20 and $v_0 \sim 10$ ms$^{-1}$ are expected.[31, 32] A value of $C = 15$ results in an estimate of $v_0/d = 1.6 \times 10^6$ s$^{-1}$. Assuming $d = 35$ μm (average core thickness for the tapes), this result gives a reasonable estimate for the flux hop velocity: $v_0 \approx 55$ ms$^{-1}$, in reasonable agreement with expectations from the literature. In comparison, the values of $C = 30$ and 40 observed here result in $v_0/d = 5.3 \times 10^{12}$ and $1.2 \times 10^{17}$ s$^{-1}$, respectively, which are both many orders of magnitude larger than physically viable.

Hence, the above experimentally derived values for $C$ of 30 and 40 are unacceptably large. These would also result in an unexpectedly strong field dependence of $C$. Consequently, we consider the standard Maley's method to yield physically unreasonable values for $C$.[23]

The results obtained after analysis with the modified form of Maley's method are shown in the Tables I, II and III for both samples and at applied fields of 0.15, 0.3, 0.5, 0.65 and 0.8 T. The analysis is limited to fields below the irreversibility field $H_{irr}(T)$. For the nonirradiated tape, $H_{irr} < 0.8$ T over a large portion of the temperature range, so there is no analysis possible for the nonirradiated tape at this applied field. Similarly, the analysis is limited to fields well above the field of full penetration $H_p$, which lies above 0.15 T in the irradiated sample for $T < 30$ K. Thus we do not include the isotherms below 30 K at this applied field for the irradiated tape.[33]

Table I contains the results found by application of the first scaling approach, utilising the temperature scaling form of Eq. (7). Since the continuity fit for each isotherm was determined by eye, values of $C$ are only accurate to ±1. A continuous $U_{eff}(J,H,T_0)$ was found for each applied field in both samples, as illustrated in Fig. 1, but the isotherms did not always fit in a smooth manner, and in some cases, the slopes of neighbouring isotherms were slightly offset. To compensate for the misalignment in one set of isotherms $C$ can be altered, but usually with the effect of bringing others out of line. Changing the form of $U(T)$ was also tried, but with similar problems. The result is the best compromise as determined by eye. Compared for an equivalent applied field and at a constant current density $J$, the effective activation energies after introduction of fission-fragment damage are approximately 50% larger than those of the nonirradiated tapes, over the entire field and current density range.

The slopes of the isotherms in $U(J)$ are much better aligned if a discrete temperature scaling factor $G$ is *chosen* for each isotherm, so that a smooth $U(J)$ is obtained. With this method, it was possible to obtain a very smooth $U(J)$ curve, with the slopes of each of the neighbouring isothermal $U(J)$ matching closely (Fig. 2). Such smooth transitions between the isothermal $U(J)$ were not obtainable with a predetermined form of $U(T)$. Consequently, we will focus our discussion on the data obtained by the empirical scaling approach.

The data set of $G$ determined at each temperature in this fitting procedure can be plotted as a function of temperature, where an empirical expression for the functional form of $U(T)$ can be fit to the data set. Figure 3 displays the empirically obtained temperature scaling set $G(T)$, along with a curve fit to the form in Eq. (9). The fitting parameters resulting from the curve fit are shown in Table II, accurate to approximately ±3%.

The empirically determined temperature scaling forms for both tapes fall into two broad groups. At low applied fields, the scaling forms are very close to the theoretically determined form $U(T) \approx (1-t)^{3/2}$ (Eq. (7)) within the specified uncertainty. Only the form of $U(T)$ at $\mu_0 H = 0.3$ T for the irradiated tape is difficult to account for, with n = -1.70 ± 0.05 being larger than the expected n = -1.50.

There is a significant change in the functional form at $\mu_0 H \geq 0.5$ T in the nonirradiated tape and similarly in fields $\mu_0 H \geq 0.65$ T in the irradiated tape. At these fields, $U(T)$ shows a remarkably different behaviour than at low fields, following a form similar to that of Eq. (8). Changes in $U(T)$ are thus more obvious when the temperature scaling term is derived directly from the data, rather than applying a term determined *a priori*.

It is generally found in the literature that a predetermined $U(T)$ is often chosen in the form of either Eq. (7) or (8).[23,27] Our empirically determined $U(T)$ for both tapes can also be fitted with these two equations. However, we would like to stress that the nature of the temperature scaling function used for the fitting of the data obtained from magnetic relaxation is determined by the type of vortex dynamics. It depends not only on fundamental pinning related parameters[23] such as $\lambda$ and $H_c$, but also on the dimensionality of the vortices, type of pinning centers, interlayer vortex coupling and many other factors. Therefore, Eqs. (7) and (8), which are derived by taking into account only the temperature dependence of $\lambda$ and $H_c$,[26] are not necessarily indicative of a fundamental description of any temperature dependencies of the magnetic relaxation data. Additionally, the successful fit of Eqs (7) and (8) to our empirically obtained $U(T)$ might only result from the fitting procedure being performed over a range of temperatures that is too narrow to discern reliably the accuracy of the fitting. Consequently, the apparently good fit of our $G(T)$ with these equations is merely coincidental and no conclusions based on their derivation can be applied to our data.

Table III contains the results for $U_{eff}$ obtained through the empirical temperature scaling approach. For the irradiated tape, $C$ is not significantly different to that determined using an *a priori* temperature scaling, listed in Table I, for applied fields $\mu_0 H \leq 0.5$ T. $C$ is increased by only slightly more than the margin of error (±1) for the nonirradiated tape in applied fields $\mu_0 H = 0.15$ and $0.3$ T. Additionally, the values determined by the empirical scaling approach for these two applied fields are very similar in both tapes, where the difference may still be due to the fitting accuracy. It indicates that $a\nu_0/d$ is very similar in both tapes at these applied fields.

In applied fields $\mu_0 H \geq 0.5$ T in the nonirradiated tape and $\mu_0 H \geq 0.65$ T in the irradiated tape, however, $C$ was determined to be significantly larger than in the lower applied fields, and also larger in comparison to the theoretically determined temperature scaling approach. This observed increase in $C$, which is correlated with the change in the form of $U(T)$, represents a change in the $\nu_0$ as the applied field is increased. The increase to $C = 19$ results in $a\nu_0/d$ less than two orders of magnitude larger than the low field estimate, which is still an acceptable estimate.

Figure 4 displays the resulting $U_{eff}(J,H,T_0)$ curves obtained using the empiric temperature scaling. A comparison at both extremes of the applied field analysed (0.65 and 0.15 T) reveals that, after introduction of fission-fragment damage, the effective activation energies are increased to approximately twice those of the control tape without fission-induced defects. This difference is at a maximum of 2.3 times for $J$ close to zero, and reduces as $J$ approaches $J_c$, with a minimum change in $U_{eff}$ of approximately 1.5 times at the upper limit of $J$ measured.

The difference in $U_{eff}$ is much more pronounced at an applied field of 0.5 T, reaching almost 15 times the nonirradiated $U_{eff}$ as $J \rightarrow 0$. This is a result of the different limits of $U_{eff}(J,H,T_0)$ as $J$ approaches zero that are observed in different applied fields and which can be better perceived by re-plotting Fig. 4 in a linear-log format, Fig. 5. In low applied fields, $U_{eff}$ diverges as $J \rightarrow 0$, as predicted by the collective creep and vortex-glass theories for an elastic FLL.[18, 19] This is distinguished from the high field behaviour, where $U_{eff}$ can be seen to approach a constant value $U_0$ as $J \rightarrow 0$, characteristic of creep by plastic shear deformations.[34]

The crossover from elastic to plastic creep is in fact correlated with the empirically determined form of $U(T)$. A diverging $U_{eff}$ is found for all applied fields where $U(T)$ is approximately described by $(1-t)^{3/2}$, whereas a form of $(1-t^2)$ defines a $U_{eff}(J,H,T_0)$ curve that approaches a constant value for $J \rightarrow 0$. In the nonirradiated tape, the crossover from divergent to non-divergent behaviour is found to occur at $\mu_0 H = 0.5$ T, but it occurs at a higher applied field, $\mu_0 H = 0.65$ T, in the irradiated tape. Therefore, at an applied field of 0.5 T, the current dependencies of the two tapes differ, and hence the diverging activation energy of the irradiated tape deviates to a large degree from the activation energy of the nonirradiated tape as $J \rightarrow 0$.

## V. DISCUSSION

The current density dependence can be better characterised by fitting the experimental $U_{eff}(J,H,T_0)$ curves with the functional form of Eq. (5) and examining the resulting values of the fitting parameter $\mu$, given in the Tables I, II and III for each different approach to the temperature scaling and plotted against applied field in Fig. 6. It can be seen that the resulting $\mu$ values are a decreasing function of applied field, with an apparent crossover from positive to negative $\mu$ values at a field $H_{cr}$. Using the empirically determined $U(T)$, this crossover occurs at $\mu_0 H_{cr} \approx 0.37$ T for the nonirradiated tape and approximately 0.65 T after irradiation. $H_{cr}$ is slightly higher when scaling is performed with a predetermined $U(T)$, but the transition of $\mu$ is evident regardless of the particular technique used for determining $U(T)$.

As noted above, changes in the divergence of $U_{eff}$, and hence the vortex dynamics, are more obvious when the temperature scaling term is derived directly from the data, rather than by applying a term *a priori*. The crossover to negative $\mu$ is then observed to occur simultaneously with the changes in the divergence of $U_{eff}$, as well as in the form of $U(T)$ and the value of $C$. These features therefore appear to be correlated.

The shift from positive to negative $\mu$ is understood to be indicative of a transition from 3D elastic creep to a 2D plastic creep regime.[35] The change in sign of $\mu$ observed thus suggests a 3D-2D transition in the system at about 0.37 T in the nonirradiated tape and $\mu_0 H \sim 0.65$ T after irradiation. In anisotropic HTS such as Bi2223, it has been established that there is a dimensional crossover in the flux lattice, when either magnetic fields or thermal fluctuations destroy the Josephson coupling between vortices in adjacent layers.[1] At low temperatures, the theoretical criterion for this field induced 3D to 2D transition is[1]

$$\mu_0 H_{cr} \propto \Phi_0/\gamma^2 s^2, \qquad (11)$$

where $\Phi_0$ is the flux quantum, $\gamma$ the anisotropy ratio, and $s$ the layer spacing. Due to a lack of exact data on the anisotropy ratio in Ag/Bi2223 samples, the theoretical crossover field can only be estimated. However, Blatter *et al.*[36] estimate $\mu_0 H_{cr} = 0.36$ T. The value of $\mu_0 H_{cr} \approx 0.37$ T observed here from the magnetisation measurements is remarkably close to this estimate.

The substantial changes observed in $U(T)$ at $\mu_0 H_{cr}$ may be consistent with a de-coupling of the pancake vortices in neighbouring layers. Clem[37] demonstrated that the inter-layer magnetic coupling between pancake vortices, in the limit of zero Josephson coupling, could be destroyed by thermal excitations even at quite low temperatures. The dimensional crossover requires a large reduction in the Josephson coupling, so that the limit of zero Josephson coupling may apply in the 2D state, above $\mu_0 H_{cr}$. Below $\mu_0 H_{cr}$, where the Josephson coupling is strong, the dependence on thermal excitations is not as strong at such low temperatures as for the 2D state. Thus, at low temperatures, the dependence of the vortex dynamics on thermal excitations will be greater in the 2D regime than in the 3D regime.

Similarly, the change in $C$ observed at $\mu_0 H_{cr}$ may be consistent with a pancake de-coupling transition. The viscosity of vortex motion depends on the pinning and interaction of the flux vortices.[36] In a crossover to 2D pancakes, the effectiveness of pinning centres on vortices in adjacent layers is reduced.[38] The weak interlayer vortex interaction also allows vortices in one layer to slip past pinned vortices in adjacent layers. This leads to a reduction in the viscosity of the vortex pancake motion. The reduced viscosity allows a faster hop velocity, which will lead to a commensurate increase in $C$.

The change in the divergent behaviour of $U_{eff}$ as $J \rightarrow 0$ may also be explained from a dimensional crossover. In the 2D state, the diverging activation energy barriers predicted for the 3D elastic FLL are not expected.[38] Above $H_{cr}$, rather, creep is dominated by the plastic shear of edge dislocations, which have a finite barrier against creep even as $J \to 0$.[36] Thus the behaviour of $U_{eff}(J,H,T_0)$ as $J \to 0$ above and below the crossover field, Fig. 5, is consistent with a dimensional crossover.

The shift in $\mu_0 H_{cr}$ to higher applied fields after irradiation signifies that the 3D vortex-glass regime is extended to higher applied fields by the fission-fragment damage centres. Hence splayed defect pinning promotes the c-axis vortex correlation.

An independent measurement of the field dependence of $\mu$ was sought to confirm the behaviour observed from the relaxation measurements. Figures 7(a) and (b) show $\ln E - \ln J$ curves at 77 K for the nonirradiated and irradiated tapes respectively. The curves have been corrected for parallel silver conduction.[39] A crossover from negative to positive curvature can be identified in both tape samples as the applied field is increased. The field at which the crossover occurs is shifted from $\mu_0 H \approx 0.08$ T in the nonirradiated tape to $\mu_0 H \approx 0.44$ T after irradiation.

As shown in Eq. (10), the change in the curvature of the $\ln E - \ln J$ curves indicates a similar change in the sign of $\mu$. The current-voltage measurements therefore also indicate a shift in $H_{cr}$ to higher applied fields as a consequence of the fission-induced damage.

The values of $H_{cr}$ derived from the magnetisation measurements, $\mu_0 H_{cr} \approx 0.37$ and 0.65 T for the nonirradiated tape and irradiated tapes respectively, are higher than those derived from the $\ln E - \ln J$ curves, $\mu_0 H_{cr} \approx 0.08$ T and 0.44 T, respectively. From fluctuation theory,[40] the temperature dependence of the crossover field, due to thermal fluctuations of the pancake vortices, is approximately given by

$$\mu_0 H_{cr} \propto 1/T. \qquad (12)$$

The magnetisation measurements are scaled to a temperature $T_0 \approx 20$ K, determined by the lowest temperature used for the scaling analysis. The $I - V$ curves, on the other hand, are measured at $T = 77$ K. Therefore, the differences in $H_{cr}$ can be explained as a result of the different measurement temperatures. We thus expect a difference of ~ ¼ between measurements of $\mu_0 H_{cr}$ at 77 and 20 K. The relative difference between the two measurements on the nonirradiated samples is ~ 0.22, in good agreement with our expectations. However, because of the strong columnar pinning, the fluctuation theory is not entirely applicable to the irradiated tape samples without alteration, where the difference is observed to be ~ 0.6. It may still account for the temperature dependence of $\mu_0 H_{cr}$, if not the magnitude of the dependence.

The $V - I$ measurements confirm that the effect of the randomly splayed quasi-columnar defect pinning is to drive the applied field at which the crossover in $\mu$ occurs to higher fields. This is consistent with the observation of the $U_{eff}(J)$ behaviour derived when the relaxation data is analysed using the *modified* form of Maley's method (see Table III and Fig. 4).

To further highlight the consistency found with the empirical temperature scaling method, we compare these results with the ones obtained in the absence of temperature scaling. The sign of $\mu$, and thus the deduced dimensionality of the system, is found to be altered radically without the application of a temperature scaling procedure. Over the entire range of applied fields, even at $\mu_0 H = 0.15$ T, $\mu < 0$ for the irradiated tape.

The results from the standard Maley's method suggest that the vortex system in the irradiated tape is in a 2D plastic creep regime over all applied magnetic fields, implying the *destruction* of the c-axis vortex coherence by the introduction of randomly splayed columnar pins and a drastic reduction in $\mu_0 H_{cr}$. The loss of vortex coherence could be explained as a result of vortex cutting[14] by the highly splayed columnar pins. This would, however, reduce the effects of the pinning to that of random point disorder,[15] inconsistent with the observed improvements in $U_{eff}(J)$, $J_c$ and the field dependence of $J_c$.[41]

Comparisons of randomly splayed quasi-columnar defects on Y123, Bi2212 and Bi2223 samples revealed greater pinning enhancements in the more highly anisotropic Bi2223.[16] At the time, the authors were unable to provide any structural evidence to explain the differences. Based on the observations obtained in this work of the increase in the dimensional crossover, and hence the c-axis vortex correlation, in fission-damaged Bi2223, it becomes clear that the difference lies in the lower anisotropy of the Y123, which promotes vortex line cutting rather than an increased inter-layer correlation.

## VI. CONCLUSION

The unphysical size and behaviour of $a v_0$ obtained in the absence of any temperature scaling, the consistency found with the empiric temperature scaling technique, and the agreement with transport voltage – current measurements reinforces the conclusion that temperature scaling is required to accurately determine the correct flux dynamics of the investigated sample. An empirically obtained temperature scaling is therefore a crucial addition to the analysis of magnetisation relaxation measurements.

In an uranium-doped, nonirradiated sample of Ag/Bi2223, we observed a transition from 3D elastic creep to 2D plastic creep regimes at about 0.37 T. Temperature scaling within the 3D system was observed to be of the form $(1-t)^{3/2}$, as predicted by theory. In the 2D system, the influence of thermal fluctuations was observed at much lower temperatures, with scaling being of the form $(1-t^2)$.

After the introduction of splayed columnar pinning centres by fission-fragment damage, we observe a large increase in the effective activation energy over a wide temperature and field range. The system remained in a 3D elastic creep regime up to higher applied fields, with $\mu_0 H_{cr} \approx 0.65$ T. The temperature scaling in the 3D and 2D states was identical to that found for the nonirradiated tape. The introduction of randomly splayed columnar defects by fission-fragment damage therefore promotes the $c$-axis vortex correlation and extends the 3D vortex-glass regime to higher applied fields. This explains the greater efficiency of the uranium-fission method in BSCCO rather than the less anisotropic YBCO.

## ACKNOWLEDGMENTS

This work was supported in part by the Australian Research Council and the Australian Institute of Nuclear Science and Engineering. The authors would like to thank Prof. H. H. Wen, Prof. H. W. Weber, and Dr J. Boldeman for helpful discussions.

**Table I:** Values of the $C$ and $\mu$ fitting parameters obtained for a fit to Eq. (5), obtained with $U(T) = (1-t)^{3/2}$.

| | 0.6-wt% UO$_4$, nonirradiated | | | | 0.6-wt% UO$_4$, $\Phi_n$ = 2.25 x 10$^{19}$ m$^{-2}$ | | | | |
|---|---|---|---|---|---|---|---|---|---|
| $H$ (T) | 0.15 | 0.3 | 0.5 | 0.65 | 0.15 | 0.3 | 0.5 | 0.65 | 0.8 |
| $C$ | 12 | 11 | 10.5 | 10.5 | 16 | 15 | 14 | 13 | 12 |
| $\mu$ | 0.3 | 0.16 | 1e-6 | -0.05 | 0.45 | 0.4 | 0.3 | 0.25 | 0.065 |

**Table II:** Fitting parameters obtained from a fit of the empirically determined data set $G(T)$ to the form in Eq. (9).

| | 0.6-wt% UO$_4$, nonirradiated | | | | 0.6-wt% UO$_4$, $\Phi_n$ = 2.25 x 10$^{19}$ m$^{-2}$ | | | | |
|---|---|---|---|---|---|---|---|---|---|
| $H$ (T) | 0.15 | 0.3 | 0.5 | 0.65 | 0.15 | 0.3 | 0.5 | 0.65 | 0.8 |
| m | 1.05 | 0.99 | 1.98 | 1.97 | 1.00 | 1.00 | 1.03 | 1.94 | 1.97 |
| n | -1.45 | -1.45 | -0.91 | -0.94 | -1.52 | -1.70 | -1.45 | -0.96 | -0.93 |
| a | 1.27 | 1.34 | 1.03 | 1.04 | 1.78 | 1.49 | 1.47 | 1.05 | 1.04 |

**Table III:** The approximate temperature scaling form $U(T)$ from the fit in Fig. 3, and the associated $C$ and $\mu$ fitting parameters.

| | 0.6-wt% UO$_4$, nonirradiated | | | | 0.6-wt% UO$_4$, $\Phi_n$ = 2.25 x 10$^{19}$ m$^{-2}$ | | | | |
|---|---|---|---|---|---|---|---|---|---|
| $H$ (T) | 0.15 | 0.3 | 0.5 | 0.65 | 0.15 | 0.3 | 0.5 | 0.65 | 0.8 |
| $C$ | 14 | 13 | 19 | 18 | 15 | 15 | 14 | 20 | 20 |
| $U(T)$ | $(1-t)^{3/2}$ | $(1-t)^{3/2}$ | $(1-t^2)$ | $(1-t^2)$ | $(1-t)^{3/2}$ | $(1-t)^{3/2}$ | $(1-t)^{3/2}$ | $(1-t^2)$ | $(1-t^2)$ |
| $\mu$ | 0.28 | 0.16 | -0.23 | -0.23 | 0.38 | 0.44 | 0.4 | -1e-6 | -0.1 |

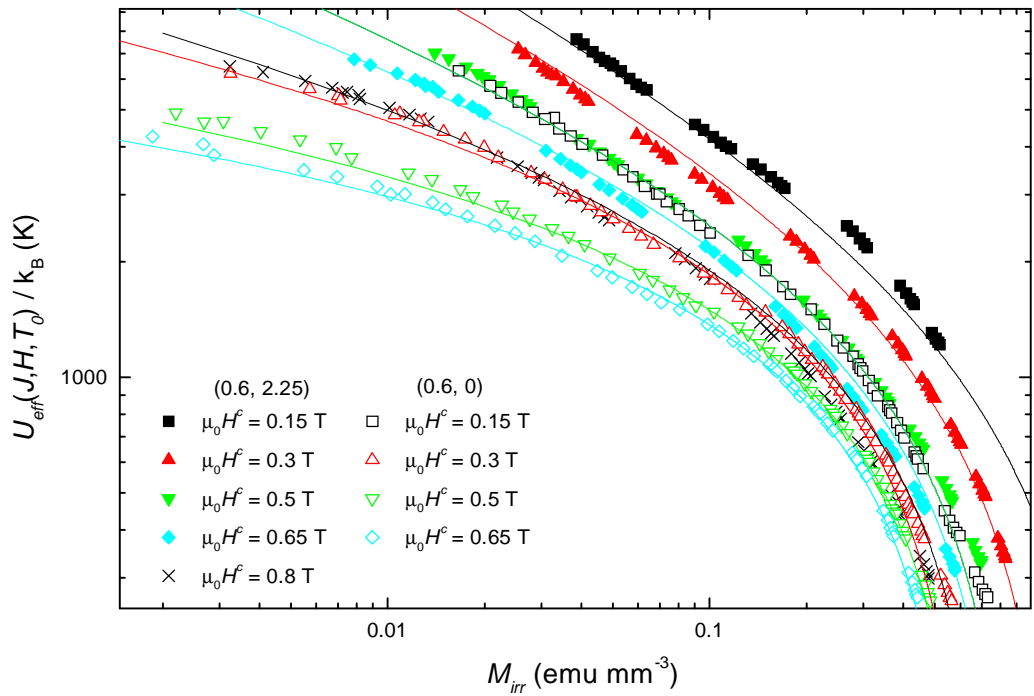

**Fig. 1.** $U_{eff}(J,H,T_0)$ for the irradiated (solid symbols) and nonirradiated (open symbols) tapes, with a theoretically determined $U(T) = (1-t)^{3/2}$. The lines are fits to Eq. (5).

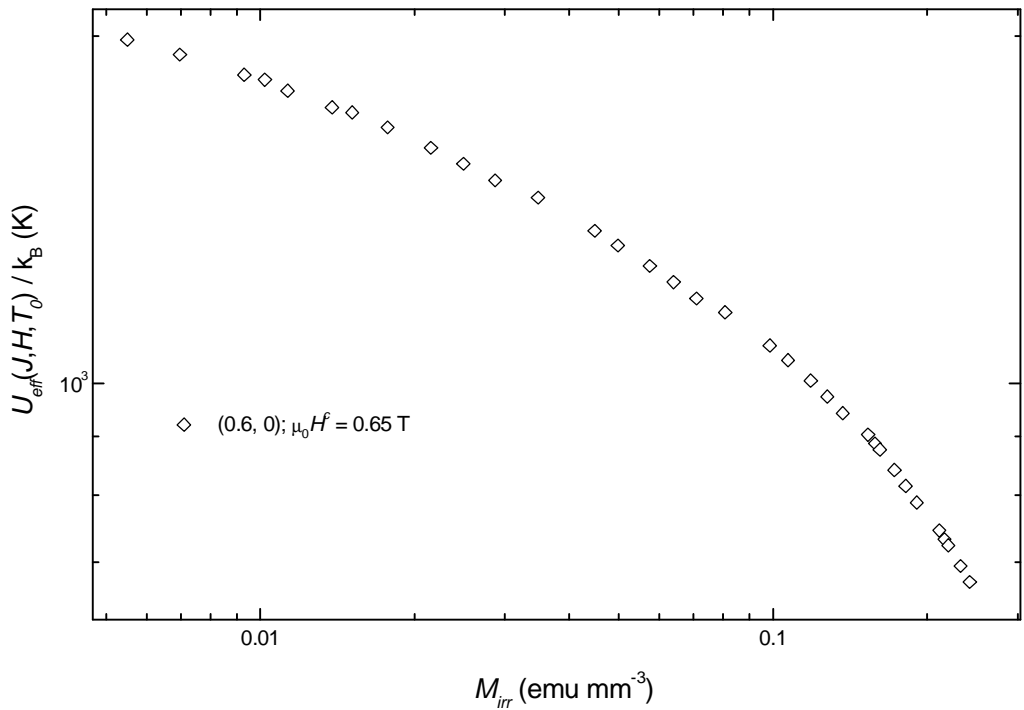

**Fig. 2.** An example of the close matching of individual isotherms using an empirical $U(T)$, showing a section of the $U_{eff}(J)$ curve obtained for the nonirradiated tape at an applied field of 0.65 T.

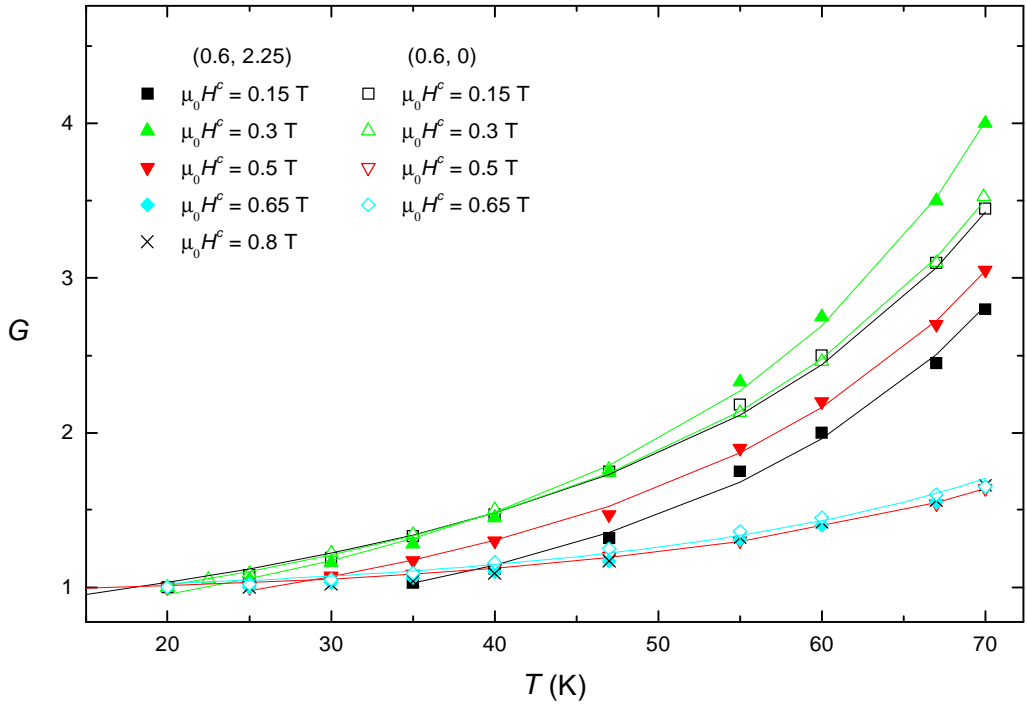

**Fig. 3.** Data set of the scaling factor *G* at each temperature, for both tapes. The errors are approximately the size of the symbols or smaller. The lines are the fits to Eq. (9).

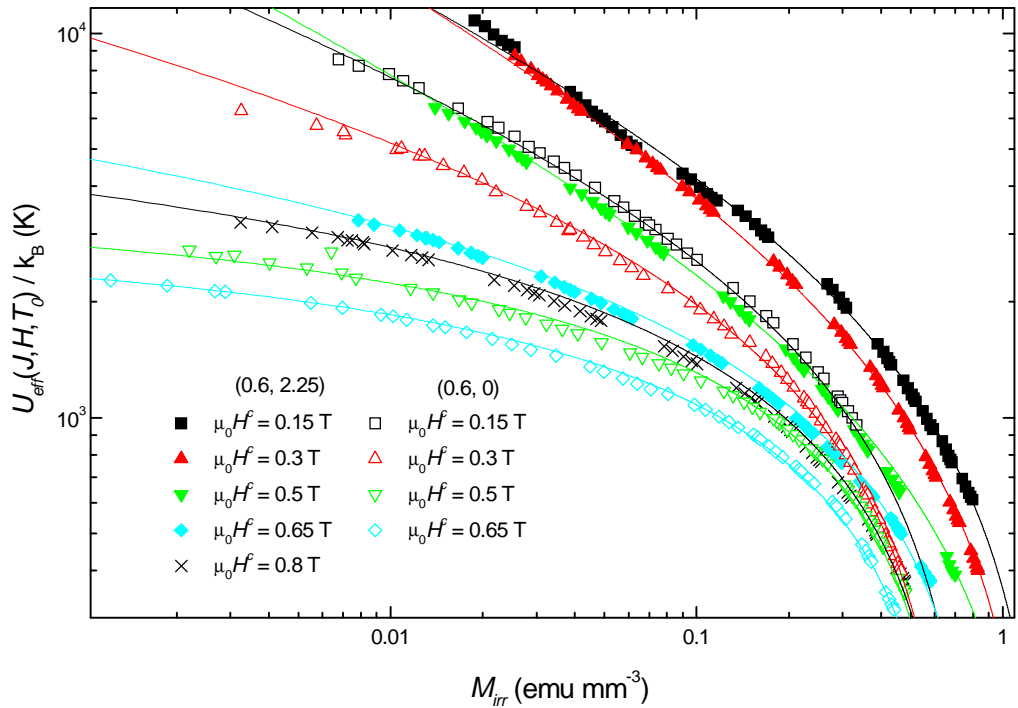

**Fig. 4.** $U_{eff}(J,H,T_0)$ for the irradiated and nonirradiated tapes with the empirically determined $U(T)$ as shown in Fig. 3. The lines are fits to Eq. (5).

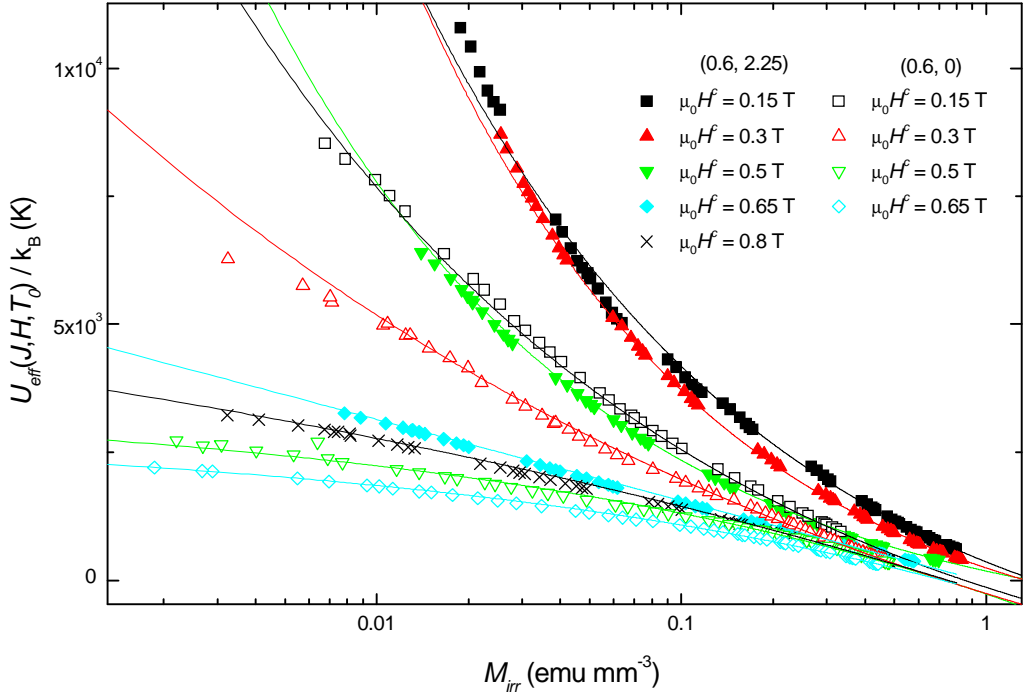

**Fig. 5.** Fig. 4 on a linear-log scale, demonstrating the different divergent behaviours of $U_{eff}(J,H,T_0)$ as $M_{irr}$ ($\propto J$) $\to 0$.

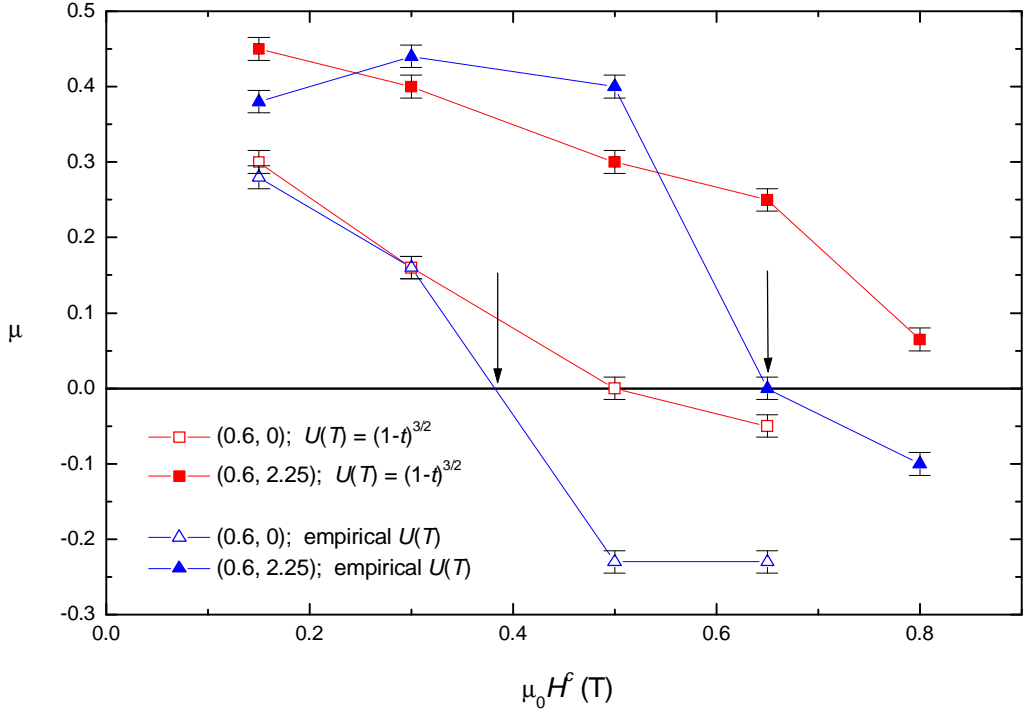

**Fig. 6.** Field dependence of $\mu$ from a fit to Eq. (5) for both the theoretically and empirically determined temperature scaling forms $U(T)$. Lines are guides to the eye only. The arrows indicate the approximate position of the crossover field $H_{cr}$.

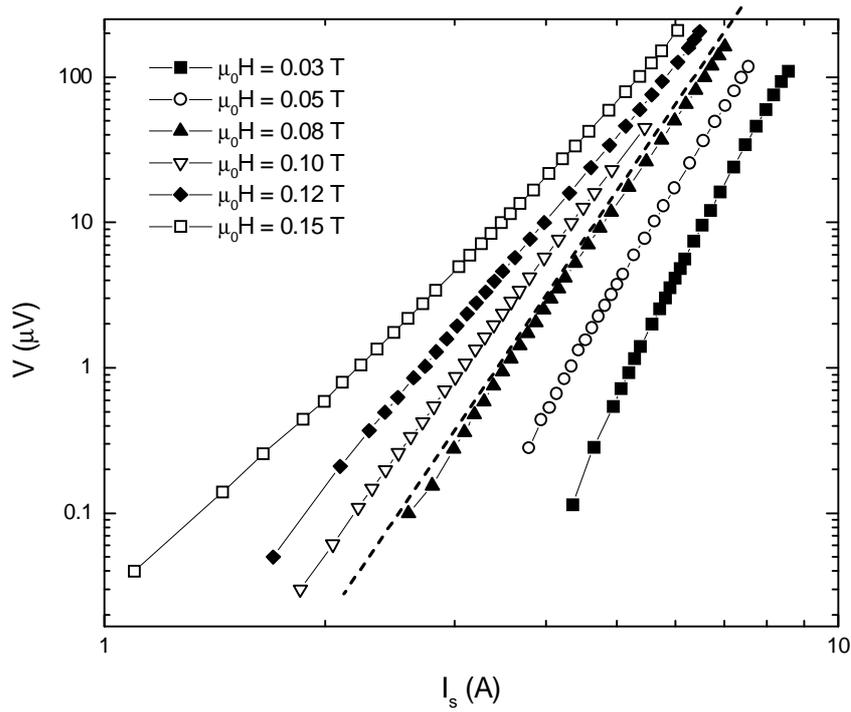

(a)

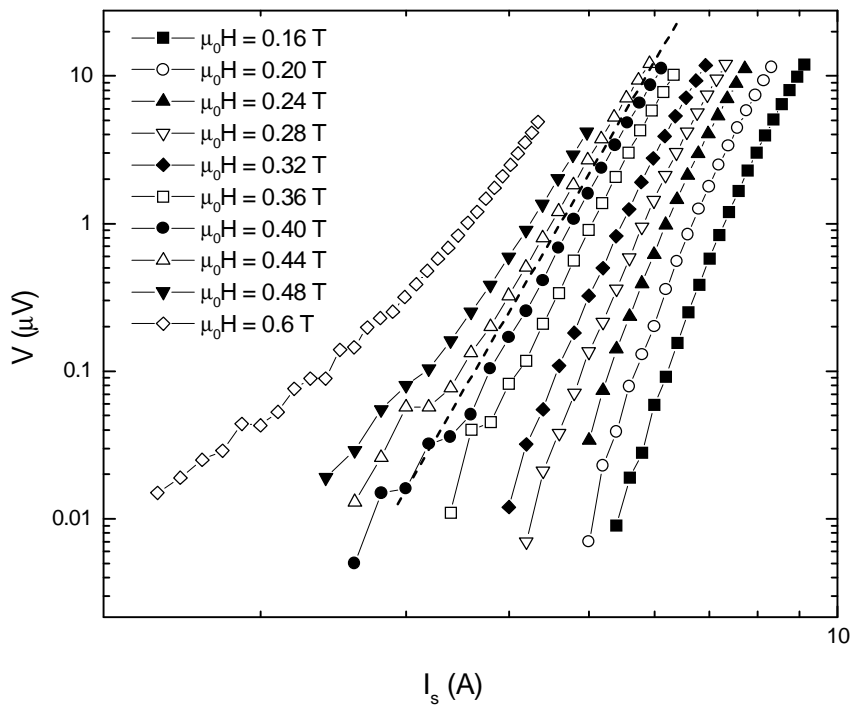

(b)

**Fig. 7.** Current – Voltage characteristics for the (a) nonirradiated and (b) irradiated tapes. Note the change in curvature evident above and below the dashed lines.